\documentclass[twocolumn,preprintnumbers,superscriptaddress,nofootinbib,aps,prl,floatfix]{revtex4}

\usepackage{footmisc,multirow}
\usepackage{subfigure,enumerate}
\usepackage{amsmath,amssymb}

\usepackage{graphicx}

\usepackage{color}

\newcommand{\be}{\begin{eqnarray*}}
\newcommand{\ee}{\end{eqnarray*}}
\newcommand{\gl}[1]{(\ref{#1})}

\newcommand{\bee}{\begin{eqnarray}}
\newcommand{\eee}{\end{eqnarray}}
\newcommand{\beeq}{\begin{equation}}
\newcommand{\eeeq}{\end{equation}}

\newcommand{\gev}{{\text{GeV}}}
\newcommand{\tev}{{\text{TeV}}}

\begin{document}

\title{Non-standard top substructure}

\begin{abstract}
  The top quark, being the heaviest particle of the Standard Model, is
  a prime candidate of where physics beyond the SM might currently
  hide before our eyes. There are many natural extensions of the SM
  that rely on top compositeness, and the top quark could follow the
  paradigm of revealing a substructure when it is probed at high
  enough momentum transfers. Observing high $p_T$ top final states
  naturally drives us towards boosted hadronic analyses that can be
  tackled efficiently with jet substructure techniques. In this paper
  we analyse the prospects of constraining exemplary non-standard QCD
  top interactions in this kinematical regime. We correctly include
  QCD modifications to additional gluon emission off the boosted top
  quark and keep track of the modified top tagging efficiencies. We
  conclude that non-standard top QCD interactions can be formidably
  constrained at the LHC 14 TeV. Experimental systematic uncertainties
  are a major obstacle of the described measurement. Unless
  significantly improved for the 14 TeV run, they will saturate the
  direct sensitivity to non-resonant BSM top physics at luminosities
  of around 100/fb.
\end{abstract}

\author{Christoph Englert} \email{christoph.englert@glasgow.ac.uk}
\affiliation{SUPA, School of Physics and Astronomy,University of
  Glasgow,\\Glasgow, G12 8QQ, United Kingdom\\[0.1cm]}

\author{Dorival Gon\c{c}alves} \email{dorival.goncalves@durham.ac.uk}
\affiliation{Institute for Particle Physics Phenomenology, Department
  of Physics,\\Durham University, DH1 3LE, United Kingdom\\[0.1cm]}

\author{Michael Spannowsky} \email{michael.spannowsky@durham.ac.uk}
\affiliation{Institute for Particle Physics Phenomenology, Department
  of Physics,\\Durham University, DH1 3LE, United Kingdom\\[0.1cm]}

\maketitle


\section{Introduction}
After the discovery of a SM Higgs boson~\cite{orig} at the
LHC~\cite{atlash,cmsh} and preliminary measurements of its properties
and couplings~\cite{atlash2,cmsh2} which indicate close resemblance to
the SM hypothesis, hints for physics beyond the SM remain elusive. A
puzzle that remains in the context of SM irrespective of a seemingly
unnatural electroweak scale is the mass hierarchy in the fermion
sector and the large mass of the top quark rather close to the Higgs
vacuum expectation value. The restoration of chiral symmetry for
vanishing Yukawa interactions guarantees that corrections to
elementary fermion masses are proportional to the fermion mass
themselves in the SM. Using the language of effective field theory,
the Yukawa couplings are marginal operators, {\it i.e.} once their
values are fixed by some UV dynamics~\cite{Froggatt:1978nt}, they
remain small at low energy scales. Hence, the large hierarchy among
the Yukawa couplings largely determined by the top quark is typically
considered a potential source of physics beyond the SM.

Indeed, the top typically plays a central role in most models that try
to explain the electroweak scale at a more fundamental
level. Supersymmetric constructions~\cite{susy}, fixed-point gravity
scenarios~\cite{fpg}, and strong interactions~\cite{Agashe:2004rs} are
just three well-known and well-established examples. In the latter
case, the large mass of the top can be understood as a (linear) mixing
effect of light elementary states with composite fermions of a
strongly interacting sector~\cite{comph,silh} that also provides a set
of Nambu Goldstone bosons forming the Higgs doublet. The mixing
effects together with fermion and gauge boson loops induce a
Coleman-Weinberg Higgs potential that triggers breaking of electroweak
symmetry at a scale much smaller than the strong interaction scale. In
such pseudo-Nambu Goldstone Higgs scenarios, we can have a large
resemblance of the Higgs phenomenology with the SM, whilst the
composite effects are hidden in the fermionic sector. Phenomenological
searches that target the potential substructure of the top quark are
therefore also extremely important in the context of Higgs physics,
since both phenomena, the ${\cal{O}}$(100 GeV) electroweak scale with
the top quark in the same ball park, might point us towards a solution
in terms of strong interactions.\footnote{It should be noted that such
  interactions typically also alter low energy observables (see, {\it
    e.g.}, Ref.~\cite{maggie}), but we remind the reader that we focus
  on the prospects of {\it direct} measurements in this work.}

\begin{figure*}[!t]
  \includegraphics[width=0.7\textwidth]{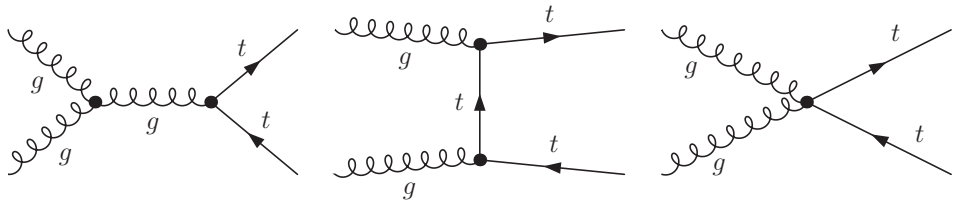}
  \caption{\label{fig:feyn} Feynman diagrams contributing to anomalous
    $p(g)p(g)\to t\bar t $ production at leading order, arising from
    the operators of Eq.~\eqref{eq:lagrangian}.}
\end{figure*}

Of course, the phenomenological implications of compositeness are not
new to particle and, more broadly speaking, to nuclear physics (see
Ref.~\cite{holst} for a review). The deviation from the anticipated
Rutherford scattering cross section at large angles observed by Geiger
and Marsden~\cite{gm} and the later resolution of atomic
nuclei~\cite{rutherford,gm2} is a well-known example of such a
programme resolving point-like sources by probing the characteristic
energy scale with high enough momentum transfers. The non-linear
structure of QCD and the mismatch of the theory's fundamental degrees
of freedom with the experimental observables, however, introduces
another layer of complexity when we deal with non-standard
interactions of a colour-charged object. We usually parametrize the
deviations from the SM via introducing higher dimensional operators in
an effective field theory description that is guided by the low-energy
gauge symmetry requirements. Since we can expect a separation between
the new physics and the electroweak scale, it is customary to
limit analyses to dimension six operator extensions to the
SM~\hbox{\cite{hikasa,wyler,Grzadkowski:2010es}}.  However, since we
cannot separate different partonic initial and final states and due to
the gauge structure, all operators that introduce non-standard QCD
properties will contribute simultaneously. Their different kinematical
dependencies can be used to disentangle
them~\cite{degrande,haberl,others,peter}, but modifications due to new
interactions will also change the response of the measurement
strategy.

The top quark production cross section will receive modifications for
energetic events if new physics in the top sector is present.  This
immediately motivates boosted top searches~\cite{boostedtops} as a
sensitive probe of modified QCD interactions on which we focus our
analysis in the following. From previous analyses~\cite{peter} it is
expected that upon correlating inclusive and boosted measurements of
$pp\to t\bar t+X$ we will be able to tightly constrain such
non-standard interactions. However, there is a caveat: top quarks when
produced at high $p_T$ are very likely to emit hard gluons before they
decay~\cite{Ferroglia:2013zwa,topradiation}. In Ref.~\cite{peter} it
was shown that such an interaction has a decreased sensitivity to
anomalous QCD top interactions. It is therefore crucial to include the
anomalous top interactions to the proper modelling of the exclusive
final state to correctly evaluate the prospects of the described
measurement. By analyzing the fully hadronized final state in such a
setup, we are also guaranteed to correctly reflect the different
selection efficiencies for the boosted subject analysis that emerge
from the BSM-induced modifications of the top spectrum.  More
precisely: we investigate the constraints that we can expect from
adapted searches for anomalous top interactions in the busy
QCD-dominated LHC environment using realistic simulation, analysis and
limit setting techniques.

Especially experimental systematics are known to be large in the tails
of top distributions where the deviations from the SM will be most
pronounced. Unless these uncertainties are properly included in the
formulation of the BSM limits we cannot trust the analysis. We discuss
the present systematics and include them to our CLs~\cite{CLS}
projection for the 14 TeV LHC run in the most conservative way. To
keep our analysis transparent we focus on two representative anomalous
top-QCD operators that are characteristic for composite fermionic
structures from a QCD point of view, namely colour charge radius and
anomalous magnetic moment~\cite{brod} (see
Ref.~\cite{schwingdeinding} for similar work on composite
leptons). The generalisation to other non-standard top-related
interactions is straightforward.

\section{A phenomenological approach to anomalous QCD top
  interactions}

To get a quantitative estimate of the leading effects of non-standard
top interactions at the LHC we focus on new physics contributions to
$t\bar{t}$ production arising from modified QCD
interactions. Non-standard electroweak properties do impact the top
decay $t\to Wb$~\cite{Bach:2012fb}, but can be studied separately in
single top-production and interlaced with our findings.

Since the current LHC searches imply strong bounds on the masses of
potential new degrees of freedom, it is expected to have a mass gap
between the SM and the BSM fields (which, {\it e.g.}, lift the top
mass via mixing effects~\cite{Grossman:1999ra}).
\begin{figure*}[!t]
  \centering
  \includegraphics[width=0.43\textwidth]{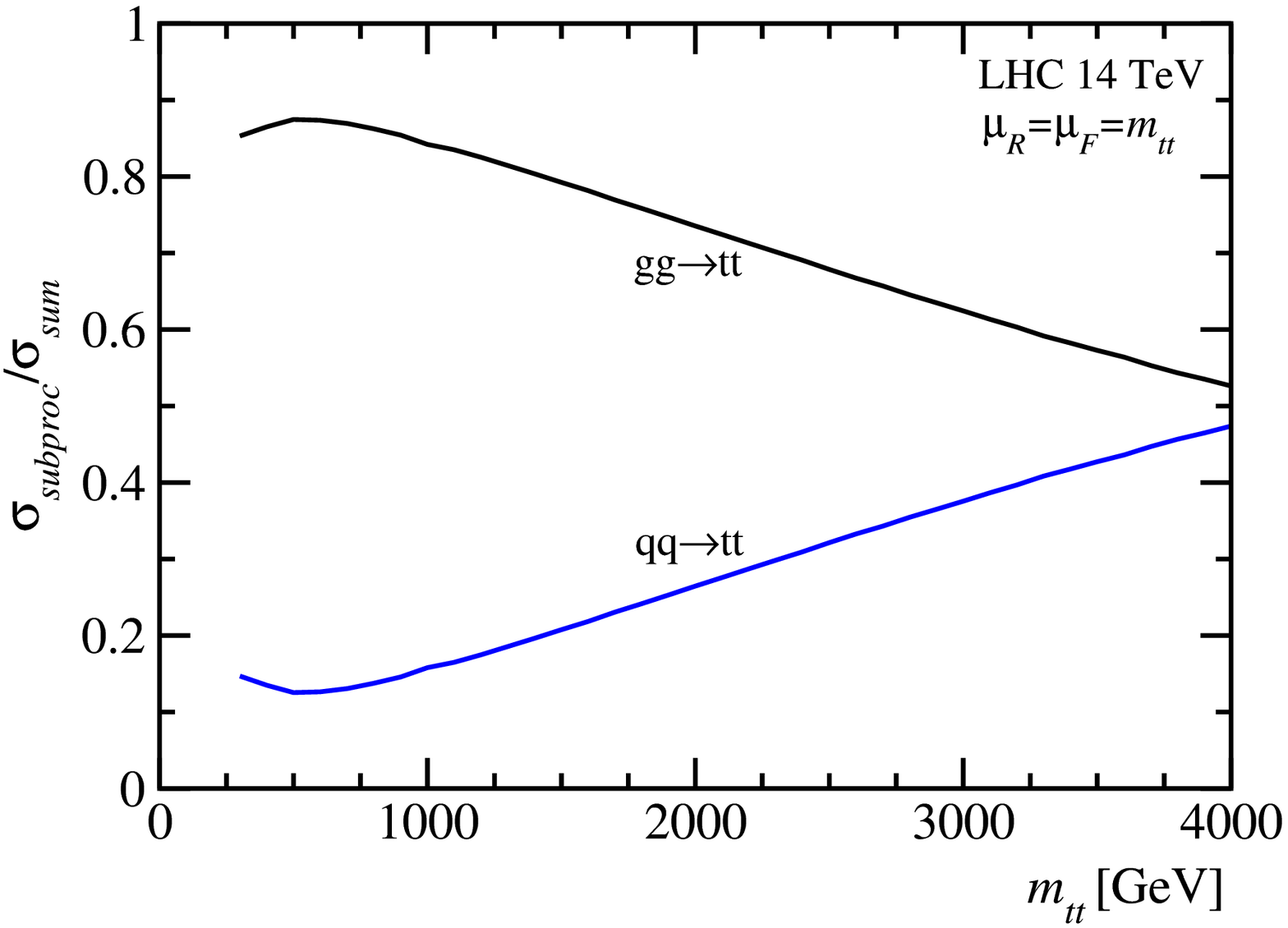}
  \hspace*{0.02\textwidth}
  \includegraphics[width=0.43\textwidth]{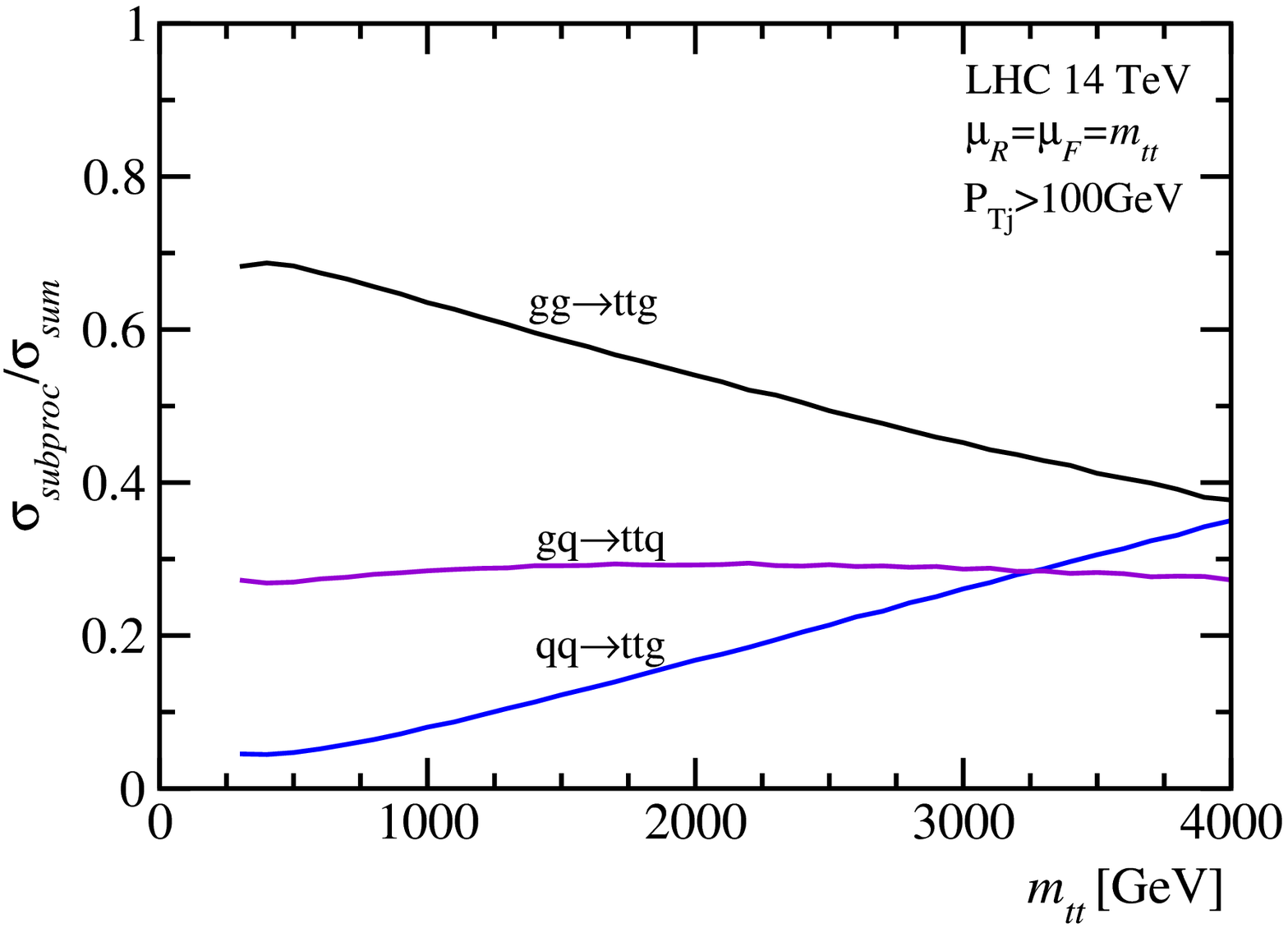} \\[0.4cm]
  \includegraphics[width=0.43\textwidth]{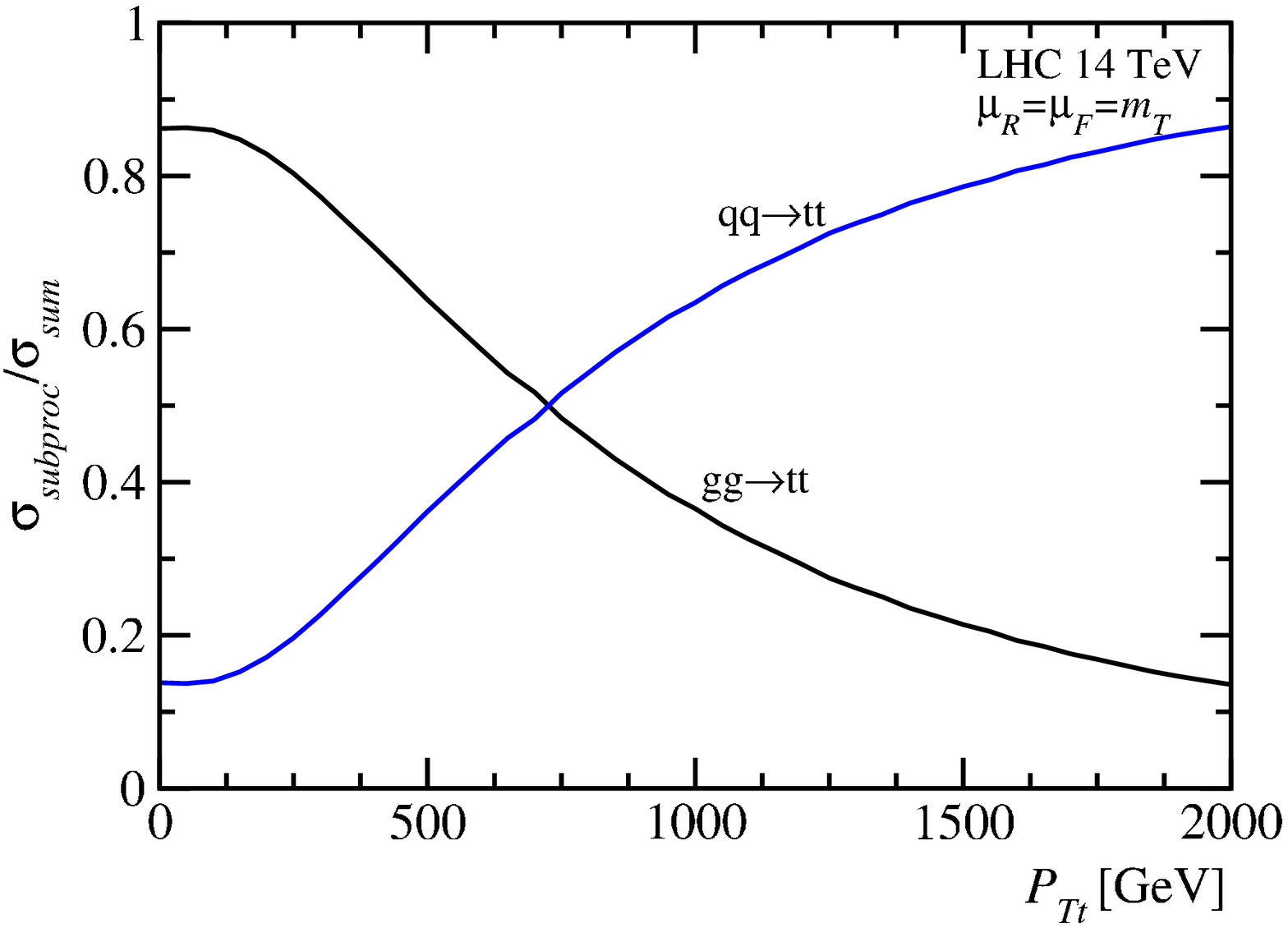}
  \hspace*{0.02\textwidth}
  \includegraphics[width=0.43\textwidth]{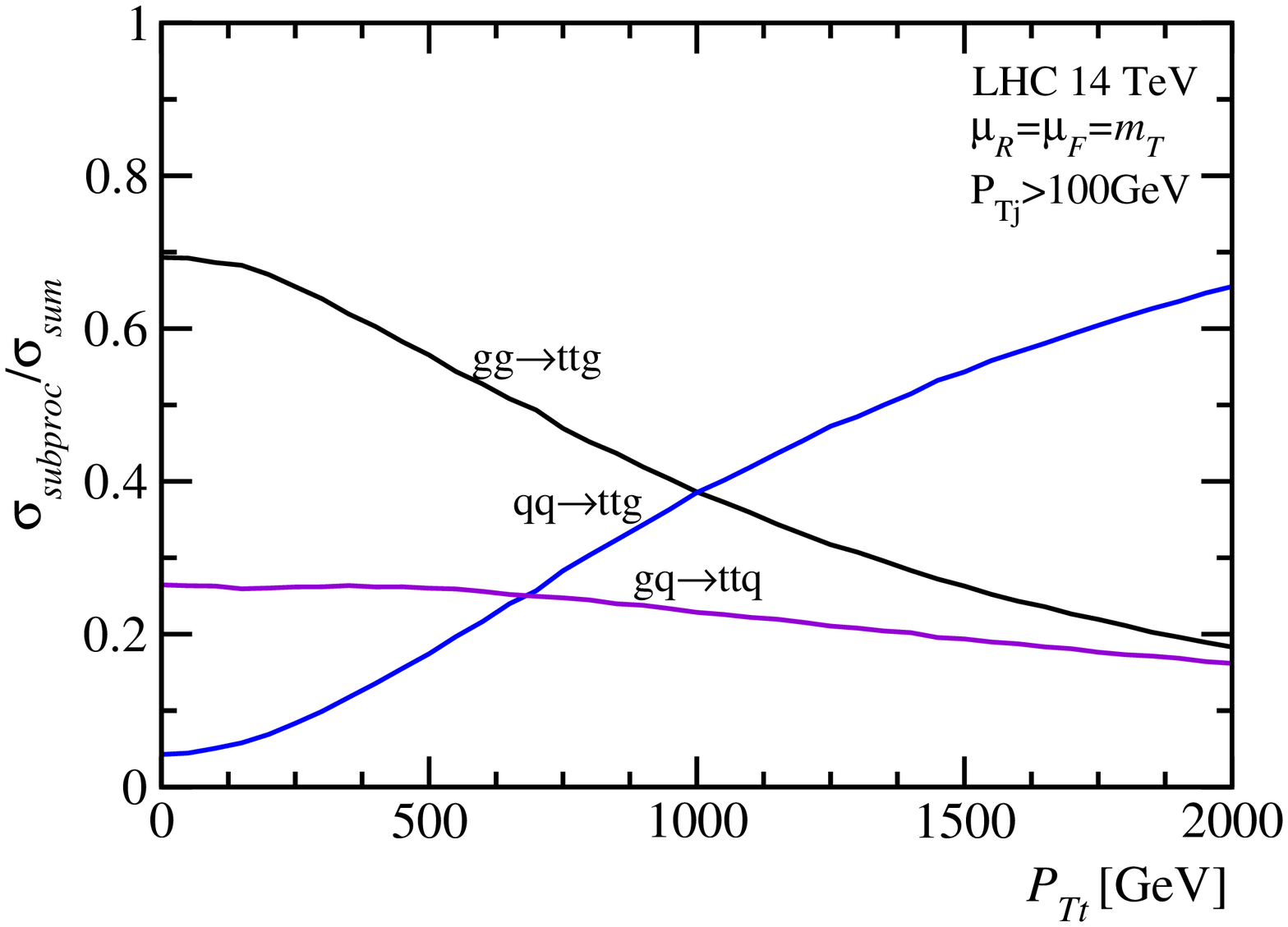}
  \caption{Fractional contribution of each partonic channel to the
    hadronic cross section for $\sigma_{t\bar t}$ (left) and
    $\sigma_{t\bar tj}$ (right) production as a function of the cut on
    the reconstructed top pair mass $m_{t\bar t}$ (top) and the transverse
    momentum of the top $p_{T,t}$ (bottom). The born cross sections
    are generated for the LHC at $\sqrt{s}=14~\tev$ with the scales
    set at the reconstructed top pair mass $m_{t\bar t}$ (top) and at the
    transverse mass $m_{T}$ (bottom).}
  \label{fig:tt_subchannels}
\end{figure*}
In this case, the new physics effects can be parametrized via higher
dimension operators involving only the SM particles and there is a
number of new contact operators which impact $t\bar t$+jets
production~\cite{wyler,degrande}. Here we focus on some operators that
allow an interpretation in terms of composite structures such as radii
and anomalous magnetic dipole moments as a proof of
principle. These non-standard properties can be introduced in a
gauge-covariant way through the following effective dimension six
interaction terms~\cite{peter,hikasa,haberl,others}
\begin{subequations}
  \label{eq:lagrangian}
  \begin{align}
    \label{eq:lagrangian1}
    \mathcal{L}_R&=
    -g_s\frac{R_t^2}{6}\bar{t}\gamma^{\mu}G_{\mu\nu}D^{\nu}t +
    {\text{h.c.}}\,, 
    \\
    \mathcal{L}_k&=
    g_s\frac{1}{4m_t}\bar{t}\sigma^{\mu\nu}(k_V+ik_A\gamma^5)G_{\mu\nu}t\,,
    \label{eq:lagrangian2}
  \end{align}
\end{subequations}
where $G_{\mu}$ is the gluon field, $G_{\mu\nu}=D_\nu G_\mu - D_\mu
G_\nu$ its field strength and $D^{\mu}=\partial^{\mu}+ig_s G^\mu$ the
covariant derivative. The convention of Eq.~\gl{eq:lagrangian} follows
Ref.~\cite{schwingdeinding}; the top quark radius $R_t$ and the
anomalous chromomagnetic and chromoelectric dipole $k_V,k_A$ moments
are related to the new physics scale $\Lambda$ in the ``traditional''
dimension six extension approach by
\begin{alignat}{5}
  R_t&=& \frac{\sqrt{6}}{\Lambda},\qquad \qquad k_{V(A)}&=&
  \rho_{V(A)} \frac{m_t^2}{\Lambda^2},
\label{eq:parameters}
\end{alignat}
where $\rho_{V(A)}$ is a $\mathcal{O}(1)$ parameter. 

To have a consistent treatment of the dimension six operator expansion
the new physics contributions are manifest only through the
interference of these new physics operators' contribution with the SM
amplitude, {\it i.e.} we do not include terms to the hadronic cross
section other than the ones that formally scale as
$\mathcal{O}(1/\Lambda^{2})$. Splitting the amplitude that results
from Eqs.~\gl{eq:lagrangian} into a SM and BSM piece
\begin{equation}
  {\cal{M}}={\cal{M}}_{\text{SM}}+{\cal{M}}_{\text{BSM}}(\Lambda^{-2})\,,
\end{equation}
we have for the (partonic) cross section 
\begin{equation}
  \label{eq:xsec}
  \sigma\sim |{\cal{M}}_{\text{SM}}|^2 + 2\Re\{
  {\cal{M}}_{\text{SM}} {\cal{M}}_{\text{BSM}}^\ast (\Lambda^{-2})\} +
  {\cal{O}}(\Lambda^{-4})\,.
\end{equation}

The expansion of the cross section to $\mathcal{O}(1/\Lambda^{2})$
removes the chromoelectric operator from the $t\bar t$
sample~\cite{haberl} and the sensitivity to $k_A$ arises from the less
dominant $t\bar tj$ contribution. The squared BSM matrix elements has
a dependence on $k_A$~\cite{haberl}. At $\mathcal{O}(1/\Lambda^{4})$,
however, when $k_A$ becomes resolvable, we can also expect additional
dimension eight operators to enter the stage via interference with the
SM amplitude. In such a case it is not clear how to interpret a limit
obtained on $k_A$. Expanding of the cross section to
$\mathcal{O}(1/\Lambda^{2})$ will therefore only yield mild
constraints on $k_A$.

\begin{figure*}[!t]
  \centering
  \includegraphics[width=0.26\textwidth, angle = 90]{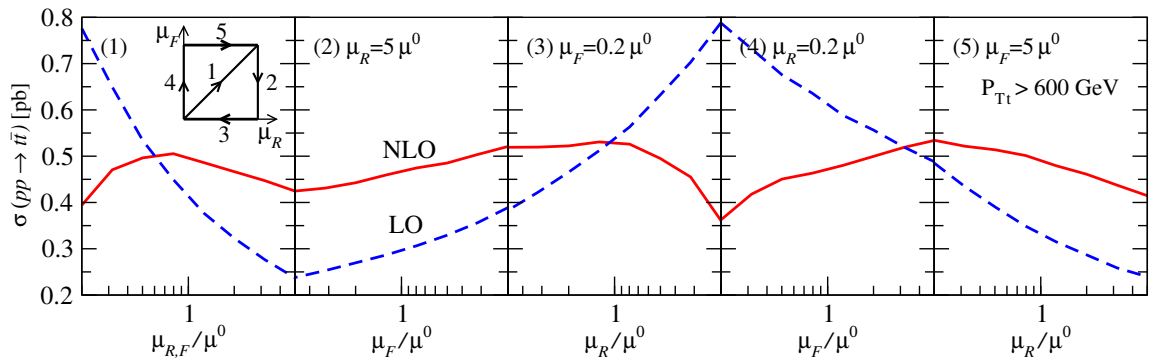}
  \caption{Renormalization and factorization scale dependencies for
    top pair production in the boosted top regime,
    $p_{T,t}>600~\gev$. The plot traces the contour in the
    $\mu_F-\mu_R$ plane with $\mu=(0.2-5)\mu^0$ as shown in the first
    panel, with $\mu^0$ defined as the event's transverse mass. The
    results are generated with {\sc{aMC@NLO}} for the LHC at
    $\sqrt{s}=14~\tev$.}
  \label{fig:scales}
\end{figure*}

The deviations $\Delta\sigma$ from the SM Born-level partonic $t\bar
t$ cross sections $\sigma_B$ sketched in Eq.~\gl{eq:xsec}
factorize~\cite{peter,degrande,haberl}:
\begin{subequations}
  \label{eq:cross_sec}
  \begin{alignat}{5}
    \frac{\Delta \sigma}{\sigma_B}(q\bar{q}\rightarrow t\bar{t})&=
    \frac{s}{3}R_t^2+\frac{6k_V}{3-\beta^2} \;, \qquad
    \label{eq:cross_sec1}\\
    \frac{\Delta \sigma}{\sigma_B}(gg\rightarrow t\bar{t}) \notag \\ &
    \hspace{-1cm} =
    \frac{k_V(36\beta-64\tanh^{-1}\beta)}{\beta(59-31\beta^2)-2(33-18\beta^2+\beta^4
      )\tanh^{-1}\beta} \;,
  \end{alignat}
\end{subequations}
where $s$ is the squared partonic center of mass energy and
$\beta=\sqrt{1-4m_t^2/s}$. Notice that for $q\bar{q}$ initial states
both new physics contributions $R_t$ and $k_V$ are present, whereas
for $gg$-induced production (the main production mode for inclusive
$t\bar{t}$ production at the LHC) there is only sensitivity to the
anomalous chromomagnetic moment $k_V$. This is due gauge invariance of the
dimension six operator, {\it i.e.}, there is a Ward identity that
guarantees the cancellation of the $R_t$ dependence\footnote{An
  identical cancellation is required to ensure a massless gluon in the
  extended theory: by closing the top-loop we have a contribution to
  the gluon two-point function from the two diagrams on the right hand
  side of Fig.~\ref{fig:feyn} which do not vanish in dimensional
  regularization.} in the sum of Fig.~\ref{fig:feyn}.  It can be shown
that for the $t\bar{t}j$ sample the same conclusion holds, {\it i.e.},
the $gg$ sub-channel still has no dependence on the $R_t$ parameter
which originates from the $q\bar{q}$ and $gq$ induced
subprocesses~\cite{peter}.

We can enhance the fraction of the $q\bar{q}$ initial state and still
probe $R_t$ at the LHC by requiring boosted top events.\footnote{A
  similar strategy has been discussed in the context of the
  central-forward top asymmetry \cite{afb}.} This is because energetic
events probe the incoming partons at high momentum fractions where the
proton's valence quarks' parton densities peak. We illustrate this in
Fig.~\ref{fig:tt_subchannels}, where we present the fractional
contribution of each partonic subprocess to the hadronic SM $t\bar
t(j) $ cross section as a function of the reconstructed $t\bar{t}$
mass and the top transverse momentum $p_{T,t}$. We can invoke cuts
on either observable to suppress the $gg$ initial state although
$p_{T,t}$ is more effective and the more crucial observable in the
context of top tagging~\cite{afb,heptoptagger}.

\section{Details, Analysis and Results}

In our analysis we focus $t\bar{t}$ production with one top decaying
semi-leptonically and the other hadronically.  As this process
involves the production of heavy coloured particles and we are
selecting the boosted kinematical regime, we can expect an important
contribution from initial and final-state jet
radiation~\cite{alwall,boosted}. To take this sufficiently into
account we include the BSM-mediated hard radiation effects via jet
merging, keeping the full BSM dependence on the non-standard
parameters of the respective samples to ${\cal{O}}(\Lambda^{-2})$.
\begin{figure*}[!t]
  \centering
  \includegraphics[width=0.43\textwidth]{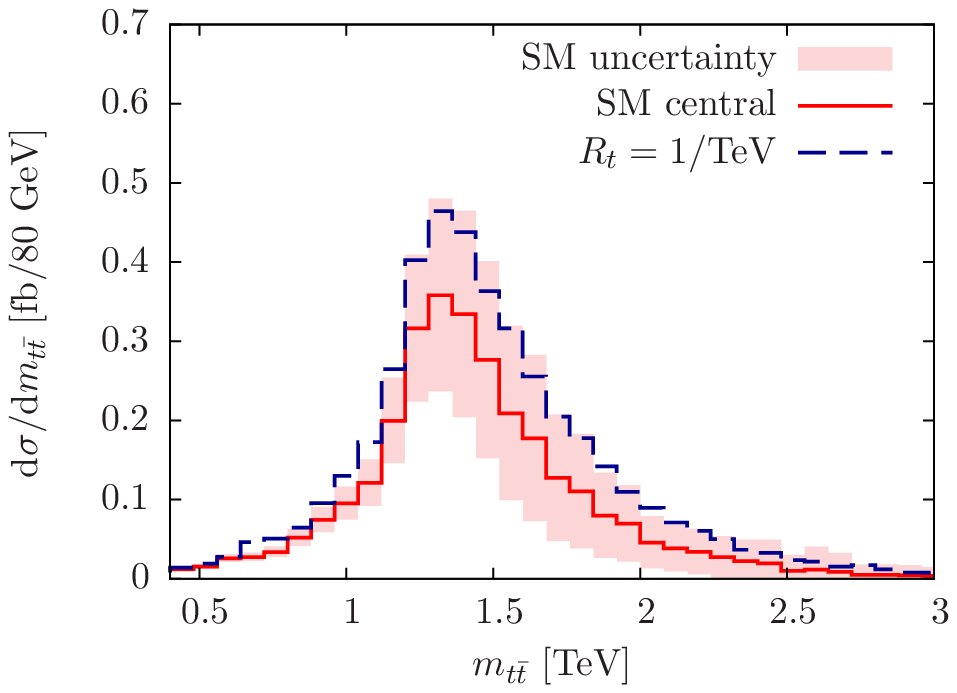}
  \hspace*{0.03\textwidth}
  \includegraphics[width=0.43\textwidth]{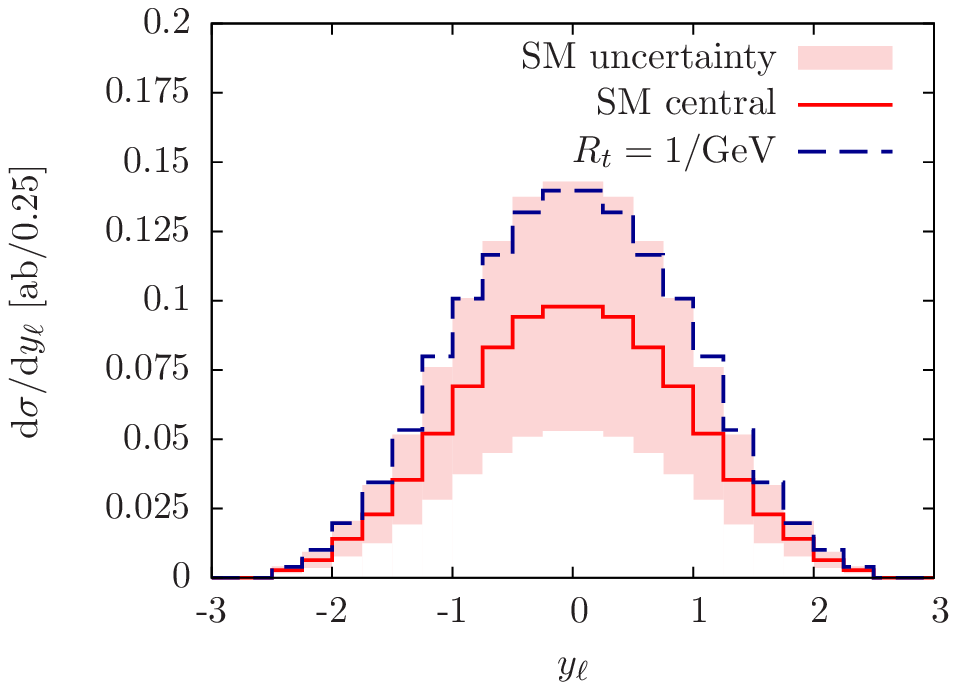} 
  \caption{Central value and uncertainty distributions of $m_{t\bar
      t}$ and $y_\ell$.  We also include an exemplary value of
    $R_t=1/{\text{TeV}}$ for comparisons.}
  \label{fig:spectra}
\end{figure*}
As already mentioned, the dependencies on the top radius arise
entirely from the $q\bar{q}$ and $qg$ initial states. Therefore, to
constrain this operator it is necessary to suppress the dominant
sub-channel at the LHC, namely the $gg$ initial state. The boosted
high $p_T$ selection serves two purposes in this sense: it removes the
less sensitive initial states and focuses on regions where deviations
from the SM are large, Eq.~\gl{eq:cross_sec}.\footnote{It is worth
  noticing that for boosted final states we do not need to worry about
  trigger issues~\cite{Aad:2012dpa,TheATLAScollaboration:2013kha}.}.

Our implementation starts by including the new interactions presented
in Eqs.~\gl{eq:lagrangian} through {\sc FeynRules}~\cite{feynrules},
which outputs a {\sc Ufo} model file~\cite{ufo} that is further used
into {\sc MadGraph5}~\cite{mg5}.  {\sc MadGraph} performs the event
generation that is subsequently showered with {\sc
  Pythia6}~\cite{pythia} where we take into account the initial and
final state radiation, hadronization and underlying event. The hard
matrix elements have been adapted to only include the interference of
the new physics amplitude with the SM counterpart; this way we
guarantee a consistent expansion of the cross section up to
${\cal{O}}(\Lambda^{-2})$ as discussed earlier when QCD emission is
hard and sensitive to the BSM effects. We have validated our parton
level matrix element implementation against existing analytic
calculations as well as an independent Monte Carlo
implementation~\cite{peter,degrande,haberl}.

The jet merging is subsequently performed by employing the MLM
scheme~\cite{mlm} as implemented in the {\sc MadGraph}
package. Throughout the analysis we consider the LHC running at
$\sqrt{s}=14~\tev$ and the SM $t\bar{t}$ cross section normalization
is re-scaled to the NNLO value,
$\sigma_{\text{NNLO}}=918$~pb~\cite{moch}.  We find that for our
boosted selection that the background is completely dominated by SM
$t\bar t$ production. All other background contributions are
negligible and well below the SM $t\bar t$ uncertainty.

We include the expected dominant NLO shape modifications via
{\sc{aMC@NLO}}~\cite{amcatnlo}: we construct a re-weighting function
with respect to the $R_t,k_V,k_A=0$ sample (the SM) to account for
differential QCD corrections in the BSM histograms. This is a
necessary procedure to have a well-defined limit $R_t,k_V,k_A\to 0$.
Throughout, we choose the renormalization
and factorization scales as the transverse mass since this choices
yield a rather flat scale dependence of the NLO matched $t\bar t$
cross section, Fig.~\ref{fig:scales}.

Instead of proceeding as in a ``traditional'' semi-leptonic $t\bar{t}$
analysis we take advantage of the efficient top tagging for high $p_T$
fat jets. This is facilitated by defining a fat jet with a large cone
size $R=1.5$ using the Cambridge/Aachen algorithm as implemented in
{\sc Fastjet}~\cite{fastjet}. We require at least one of these objects
to have a transverse momentum larger than
$p_{T,\text{fatjet}}>600~\gev$. We choose this exemplary value due to
a large top tagging efficiency $\sim 30\%$ and small fake rate $\sim
3\%$. For this threshold the $t\bar t$ cross section is also still
large enough ${\cal{O}}\text{(pb)}$ to perform measurements with small
statistical uncertainties; the eventual value of $p_{T,\text{fatjet}}$
by the experiments will optimise the systematic uncertainty. This fat
jet is then further processed by the {\sc
  HEPTopTagger}~\cite{heptoptagger}. Initially the {\sc HEPTopTagger}
was designed to reconstruct only mildly boosted top quarks ($p_{T,t}
\simeq m_t$) using a very large fat jet cone size. However, in
searches for heavy resonances~\cite{heptopexp} it was shown that due
to its flexible reconstruction algorithm and jet grooming procedures
the {\sc HEPTopTagger} is an effective tool to reconstruct highly
boosted top quarks while maintaining a small background fake
rate. Other top taggers, designed to tag highly-boosted top quarks,
can be similarly effective \cite{othertoptaggers, topradiation}.  Top tagging is
sensitive to the top's $p_T$ BSM spectrum modification and modified
hard shower profile that results from including $t\bar t j$ at
${\cal{O}}(\Lambda^{-2})$ precision. Hence, the top tag efficiency
itself is a function of the anomalous parameters.
 
After a successful tag, the corresponding jet is removed from the
event and we proceed by re-clustering the remaining hadronic activity
as usual, {\it i.e.} by applying the Cambridge/Aachen algorithm with
$R=0.5$. Jets are selected with properties $p_{T,j}>30~\gev$ and
$|\eta_j|<4$. We also require an isolated lepton in the final state
with $p_{T,\ell}>20~\gev$ and $|\eta_\ell|<2.5$ where the lepton is
defined isolated if the transverse energy deposit $E_{T,\text{had}}$
inside a cone around the lepton of size $R=0.2$ is less than 20\% of
its transverse energy $E_{T,\ell}$.

On the one hand, the small theoretical uncertainties on the $t\bar{t}$
invariant mass motivates this observable as a suitable choice to
examine our BSM hypotheses~\cite{Ferroglia:2013zwa}. From
Eq.~\gl{eq:cross_sec} it becomes clear that dominant BSM corrections
are directly reflected in the $m_{t \bar t}$ distributions (it is also
the variable which typically enters as the only kinematical parameter
in total cross section and re-summation calculations,
see~\cite{Ferroglia:2013zwa,moch}). On the other hand, the transverse
fat jet momentum and lepton pseudorapidity $y_\ell$ determine the
$t\bar t+\hbox{jets}$ kinematics to a large extent for boosted final
states. From a boosted top reconstruction point of view,
$p_{T,\text{fatjet}}$ is the crucial observable as the threshold
largely determines the working point. Since we choose a specific value
for $p_{T,\text{fatjet}}$ in our analysis, we turn to $m_{t\bar t }$
and $y_\ell$ in the following.

\begin{figure*}[!t]
  \centering
  \parbox{0.44\textwidth}{
    \subfigure[][\label{fig:rt}]{
      \includegraphics[width=0.43\textwidth]{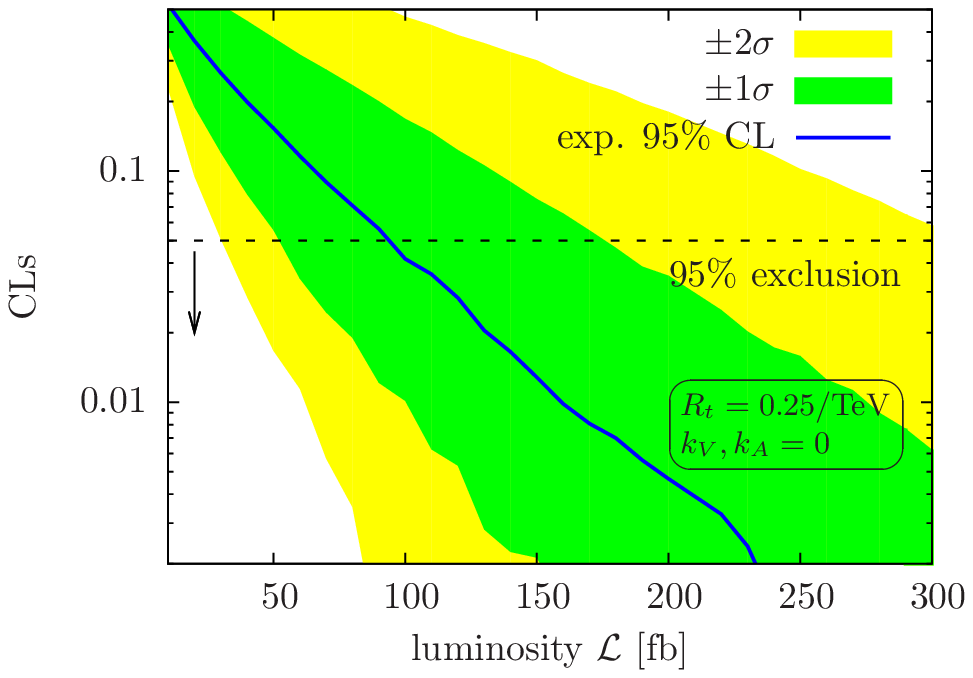}
    }
  }\hspace*{0.02\textwidth}
  \parbox{0.44\textwidth}{
    \subfigure[][\label{fig:kv}]{
      \includegraphics[width=0.43\textwidth]{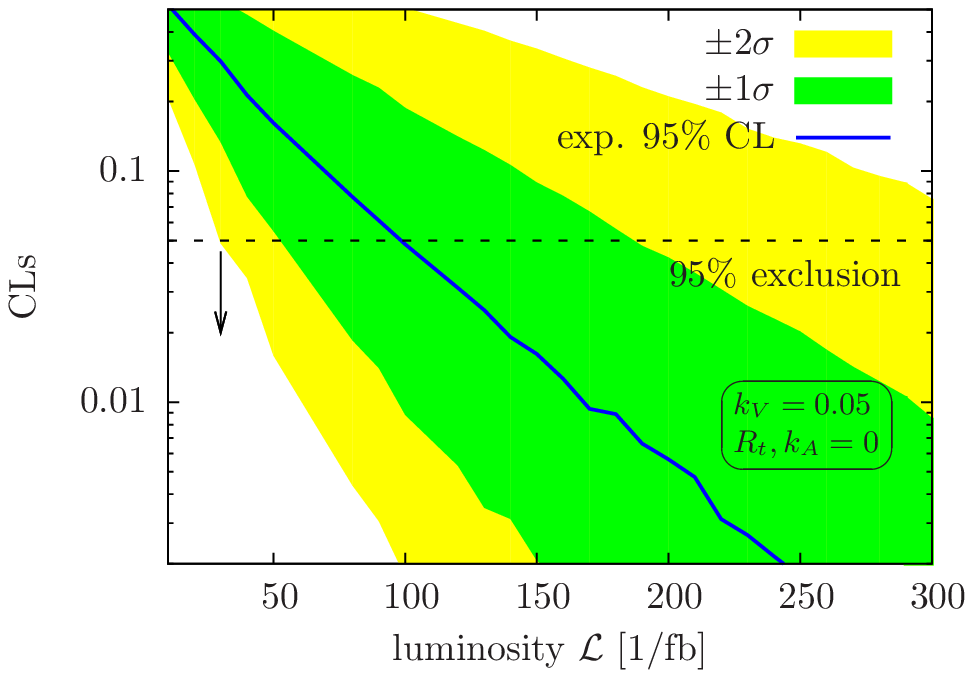}
    }
  }
  \caption{Confidence level contours for the operator
    Eq.~\gl{eq:lagrangian} in a boosted analysis of $pp\to t\bar
    t+\text{jets}$ for 14 TeV collisions as described in the text. We
    pick values of $R_t,k_V$ that can be constrained at luminosities
    of around 100/fb close to the systematics' threshold.}
    \label{fig:lag1cls}
\end{figure*}

\begin{figure}[!b]
  \includegraphics[width=0.43\textwidth]{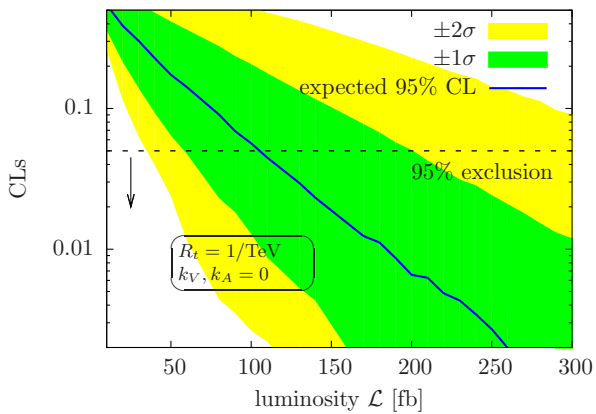} 
  \caption{Confidence level contours for the operator
    Eq.~\gl{eq:lagrangian1} in a boosted analysis of $pp\to t\bar
    t+\text{jets}$ for 14 TeV collisions as described in the text for
    a value of $R_t=1/\text{TeV}$ based on $y_\ell$. Choosing
    $m_{t\bar t}$ as discriminant results in a factor $\sim 4 $
    improvement of the limit setting, Fig.~\ref{fig:lag1cls}.}
    \label{fig:lag2cls}
  \end{figure}

Missing energy of the final state from the leptonic top decay is not a
drawback: the final state neutrino momentum can be reconstructed by
requiring transverse momentum conservation and by imposing that the
invariant mass $\ell^{\pm}$--neutrino is equal to $m_W$. These
conditions define respectively the neutrino transverse and
longitudinal momentum components. To suppress the combinatorics in the
$t\bar{t}$ mass reconstruction we need to identify which jet is the
most likely to be the $b$-jet, despite of not using $b$-tagging in
this analysis. This can be efficiently done by identifying the $b$-jet
as the closest jet to the lepton with an invariant bottom-lepton mass
that satisfies the top decay kinematics~\cite{Plehn:2011tf}
\begin{equation}
  m_{b\ell} <
  \sqrt{m_t^2-m_W^2}\simeq 154.6~\gev\,.
\end{equation}

After these steps we end up with distributions as depicted in
Fig.~\ref{fig:spectra}; the BSM-induced shape
modification includes a lot of information that we would like to
exploit in a binned hypothesis test based on sampling the
log-likelihood
\begin{multline}
  \label{eq:like}
  {\cal{Q}}= -2 \sum_{i\in \text{bins}} n_i^{\text{pseudo}}\log
  \left(1+{n_i^{\text{BSM}}\over n_i^{\text{SM}}}\right) -
  \text{const}
\end{multline}
with Monte Carlo pseudo-data $\{n_i^{\text{pseudo}}\}$, given the
input of the (B)SM histograms
$\{n_i^{\text{(B)SM}}\}$~\cite{CLS,junk}.

There is a caveat. The uncertainties, especially in the $m_{t \bar t
}$ tails of the distributions can be large, and are currently driven
by experimental systematics~\cite{Aad:2012dpa} rather than theoretical
limitations (for a recent high precision calculation
see~\cite{Ferroglia:2013zwa}). To get a feeling of the size of the
systematics we include the relative systematic uncertainty
from~\cite{Aad:2012dpa} for $\sqrt{s}=7$~TeV to
Fig.~\ref{fig:spectra}; the theoretical uncertainty
of~\cite{Ferroglia:2013zwa} is negligible compared to the systematics
of~\cite{Aad:2012dpa}. We map the integrated $m_{t \bar t}$
uncertainty to a flat $y_\ell$ uncertainty; for central tops at
transverse momenta of the order of 600 GeV this is a reasonable
approximation. It becomes immediately clear that the shape uncertainty
will be the limiting factor of this analysis, especially if we want
push limits $R_t,k_V,k_A\to 0$.

The standard way of including such an uncertainty is via nuisance
parameters of the null hypothesis (SM $t\bar t+\hbox{jets}$ production
in our case)~\cite{CLS,junk,cranmer}. When computing the confidence
level, these nuisance parameters are marginalized or
profiled. However, it can happen that the process of marginalization
can stealth the systematic uncertainty entirely. By, {\it e.g.},
including a shape uncertainty to only the null hypothesis and not to
the alternative hypothesis, marginalization will shift the median of
the toy-sampled log-likelihood distribution for the null hypothesis
away from the alternative hypothesis' median.  The exclusion in
this case appears to be larger than it should be, especially when the
uncertainty bands overlap with the difference of null- and alternative
hypothesis. To avoid issues of this type we include {\it only} bins
which exceed the SM uncertainty to the log-likelihood; {\it i.e.} our
null hypothesis is the one sigma upwards fluctuated SM
hypothesis. This way we reflect the systematic uncertainty in an
extremely conservative way; profiling or marginalization will
correctly reduce the uncertainty when correlations with other signal
regions ({\it e.g.} total cross sections and subsidiary top
measurements using the ABCD method) are taken into account. This is
information which requires access to the LHC data samples is not
available to us and also somewhat beyond the scope of this work. We
remind the reader to keep in mind that the outlined analysis when
performed by the experiments is likely to yield improved constraints
eventually.

From Eq.~\gl{eq:like} it is clear that the binned log-likelihood
approach will pick up sensitivity from regions in the single-valued
discriminant where $n^{\text{BSM}}_i/n^{\text{SM}}_i$ is large but
still resolvable according to our definition. Hence, the sensitivity
is dominated by the $p_T$ threshold behavior of the $t\bar t $ sample
and jet radiation. There the uncertainty is comparably low $\sim 20\%$
and the {\it absolute} cross section modification large (keep in mind
that the tails of the parton-level distribution grow according to
Eq.~\gl{eq:cross_sec}, which does not include the pdf suppression,
which quickly limits the considered analysis statistically).

\begin{figure}[!t]
  \includegraphics[width=0.43\textwidth]{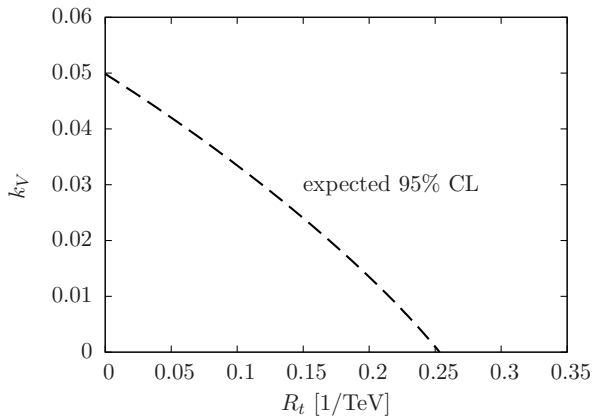} 
  \caption{Confidence level contour for operator
    Eq.~\gl{eq:lagrangian} in a boosted analysis of $pp\to t\bar
    t+\text{jets}$ for 14 TeV collisions as a function of
    $R_t,k_V,k_A=0$.}
    \label{fig:lag2cls2d}
  \end{figure}

We show the expected 95\% exclusion as a function of the integrated
luminosity $\cal{L}$ in Fig.~\ref{fig:lag1cls} for three different
samples that can be excluded with a data sample of 100/fb at a 14 TeV
LHC. The width of the 1 and 2 sigma bands being rather large indicates
that we are very close to the border of the discriminable parameter
region (in terms of our definition laid out in the previous
section). Indeed, for smaller individual values $R_t,k_V$ we cannot
formulate constraints as the BSM distribution is entirely covered by
the SM uncertainty band. We therefore conclude that an improvement
beyond the shown parameter choices depends crucially on the reduction
of the experimental systematics (which should be well-possible when
larger data samples are available). As expected the expected
constraints from using $m_{t \bar t}$ as a single discriminant are
superior to integrated sensitivity observables such as $y_\ell$,
Fig.~\ref{fig:lag2cls}.

Comparing to the preliminary investigations of Ref.~\cite{peter}, we
find that applying statistical algorithms as applied by the
experiments and realistic simulation and analysis approaches, we find
constraints in roughly the same parameter region: $R_t\lesssim
0.25/\tev$ and $k_V\lesssim 0.05$ at 95\% CL. And extrapolation into
the $(R_t,k_V,k_A=0)$ plane is shown in
Fig.~\ref{fig:lag2cls2d}. Since we include a differential shape
information of the top spectrum and a lower $p_T$ threshold that
guarantees a quick saturation of the statistical uncertainty at
comparably small luminosities we obtain more stringent expected
constraints than simple correlations of inclusive and exclusive
measurements, even when the systematic uncertainty is larger. Working
in a consistent expansion to $\sim\Lambda^{-2}$, we can only obtain
unrealistically large values on $k_A \gg 1$ that feed into our results
through higher jet multiplicities exclusively.\footnote{Going beyond
  the $\sim\Lambda^{-2}$ approximation will be unavoidable if an
  excess in the tail will be observed with the described limit-setting
  analysis that implements a practitioners' approach.}

\section{Conclusions and Outlook}

After the discovery of a Higgs boson that seems to follow the
SM-paradigm and the lack of any hints towards natural physics
completions at the TeV scale prompts us study the heavy degrees of
freedom of the SM more carefully. Top quark physics, typically
considered an impediment for new physics searches by providing a major
background contribution, is a well-motivated candidate for such
analyses. On the one hand, the properties of the top quark are still
largely unknown, even after it was discovered nearly twenty years
ago. On the other hand, the abundant production of top pairs at the
LHC allows us to tightly constrain smallest resolvable deviations from
the SM-predicted coupling pattern that is expected to be observed if
the top quark arises (partially) as a bound state of a strongly
interacting sector. This option is widely discussed in the literature
and investigating anomalous QCD interactions in the top sector
provides a path to either observe our strongly constrain such a
scenario.

Resolving a potential composite structure with large momentum
transfers in the top sector naturally motivates boosted top analyses
as highly sensitive channels. Reconstruction techniques are under good
theoretical control and have successfully been applied in $t\bar t$
resonance searches~\cite{Aad:2012dpa}. Such resonances are expected in
strongly interacting theories, too, but typical composite interactions
can be expected to predominantly manifest themselves in a large
deviation of the $t\bar t$ spectrum's tail and experimental and
theoretical uncertainties become major limitations of such searches.

In this paper we have computed the expected 95\% confidence level
constraints on a set of non-SM effective top QCD interactions
resulting from an exemplary boosted top analyses and a representative
set of operators. We have included the dominant first hard gluon
radiation effects in a matched approach. Systematic differential
uncertainties are taken into account in the most conservative way, and
are based on current 7~TeV measurements. We therefore expect our
constraints to be on the conservative end and believe that the actual
analysis when performed by the experiments can indeed improve on our
results.

Our hadron-level analysis correctly captures the top tagging's varying
efficiency as function of the anomalous parameters. This together with
a state-of-the-art binned log-likelihood formulation of the expected
confidence level constraints shows that differential shape information
supersedes the naive extrapolation of earlier theoretical work, even
when errors are considerably larger. We find that we should be able to
probe an anomalous chromomagnetic moment at the per cent level and
QCD-induced top radii at $\lesssim 0.25/\tev$.

In summary, the search for a potential top substructure strongly
benefits from recent developments in jet substructure analysis
techniques. Adapting existing boosted top searches to BSM scenarios of
this type is a straightforward exercise in the light of the results of
Ref.~\cite{Aad:2012dpa}. Given that this is an alternative route to
study theoretically well-motivated scenarios beyond the SM we hope
that this is incentive enough for the experiments to eventually
perform measurements as outlined here.

\subsubsection{Acknowledgments} 
We thank James Ferrando and Olivier Mattelaer for helpful
conversations. We also thank Ben Pecjak for providing the results of
Ref.~\cite{Ferroglia:2013zwa}. CE thanks David Miller, Liam Moore,
Michael Russell, and Chris White for discussions on the topic. CE is
supported in parts by the IPPP Associateship programme.


\end{document}